\documentclass[]{interact}

\usepackage{subfigure}
\usepackage{booktabs,multirow}
\usepackage{natbib}
\bibpunct[, ]{(}{)}{;}{a}{}{,}
\usepackage{hyperref}
\def\equationautorefname~#1\null{Equation~(#1)\null}
\hypersetup{
  colorlinks=true,
  linkcolor=blue,
  citecolor=blue,
  urlcolor=blue}
\usepackage[labelsep=period]{caption}

\let\proglang=\textsf
\newcommand{\pkg}[1]{{\fontseries{b}\selectfont #1}}

\theoremstyle{plain}

\theoremstyle{definition}

\theoremstyle{remark}

\usepackage{makecell}
\usepackage{todonotes}
\usepackage{xcolor}

\usepackage{setspace}
\begin{document}


\title{Feature-based intermittent demand forecast combinations: accuracy and inventory implications}

\author{
\name{Li Li\textsuperscript{a}, Yanfei Kang\textsuperscript{a}, Fotios Petropoulos\textsuperscript{b} and Feng Li\textsuperscript{c}\thanks{CONTACT Feng Li. Email: feng.li@cufe.edu.cn }}
\affil{\textsuperscript{a}School of Economics and Management, Beihang University, Beijing, China; \textsuperscript{b}School of Management, University of Bath, UK;
	\textsuperscript{c}School of Statistics and Mathematics, Central University of Finance and Economics, Beijing, China}
}

\maketitle

\begin{abstract}
Intermittent demand forecasting is a ubiquitous and challenging problem in production systems and supply chain management. In recent years, there has been a growing focus on developing forecasting approaches for intermittent demand from academic and practical perspectives.  However, limited attention has been given to forecast combination methods, which have achieved competitive performance in forecasting fast-moving time series.  The current study aims to examine the empirical outcomes of some existing forecast combination methods and propose a generalized feature-based framework for intermittent demand forecasting.
The proposed framework has been shown to improve the accuracy of point and quantile forecasts based on two real data sets. Further, some analysis of features, forecasting pools and computational efficiency is also provided. The findings indicate the intelligibility and flexibility of the proposed approach in intermittent demand forecasting and offer insights regarding inventory decisions.

\end{abstract}


\newpage
\setcounter{page}{1}

\section{Introduction}

\label{sec:intro}

Intermittent demand with several periods of zero demand is ubiquitous in practice.  Over half of inventory consists of spare parts, in which demand is typically intermittent \citep{nikolopoulos2021we}. Given the high purchase and shortage costs associated with intermittent demand applications, accurate forecasts could be coupled with improved inventory management in the field of manufacturing \citep{jiang2020intermittent}, aerospace \citep{wang2016select}, retailing \citep{sillanpaa2018forecasting} and so on \citep{balugani2019periodic,babai2019new}.

What makes intermittent demand challenging to forecast is that there are two sources of uncertainty: the sporadic demand occurrence, and the demand arrival timing. Seminal work on intermittent demand forecasting by \cite{croston1972forecasting} proposed to forecast the sizes of demand and the inter-demand intervals separately.  Then some scholars followed this idea and put forward some developments. For example, Syntetos-Boylan Approximation (SBA) proposed by \citet{syntetos2005accuracy} delivered approximately unbiased estimates and constituted the benchmark in subsequently proposed methodologies for intermittent demand forecasting.

\cite{syntetos2005categorization} proposed a
 categorization of demand patterns to facilitate the selection of \citet{croston1972forecasting}'s method and SBA \citep{syntetos2005accuracy}. A classification rule was expressed in terms of the average inter-demand interval and the squared coefficient of variation of demand sizes \citep{syntetos2005categorization}. \cite{kostenko2006note} developed the SBC categorization scheme \citep{syntetos2005categorization} and
 suggested a simple and more accurate rule, which has been widely used in the research of intermittent demand \citep{petropoulos2015forecast, spiliotis2021product}.

However, \citet{croston1972forecasting}'s method and SBA update demand sizes and intervals, which leads to inapplicability in periods of zero demand when considering inventory obsolescence. To overcome this shortcoming, \citet{teunter2011intermittent} proposed a new method called Teunter-Syntetos-Babai (TSB) to update the demand probability instead of the demand interval. TSB has been proved to have good empirical performance for the demands within linear and sudden obsolescence \citep{babai2014intermittent}.

The aforementioned forecasting methods for intermittent demand are all parametric methods, which estimate the parameters of a specific distribution. Instead, non-parametric intermittent demand methods directly estimate empirical distribution based on past data, with no need for any assumption of a standard probability distribution. The bootstrapping methods, and the overlapping and non-overlapping aggregation methods dominate the research field of non-parametric intermittent demand forecasting \citep{willemain2004new, hasni2019performance, hasni2019spare, boylan2021intermittent,boylan2016performance}.

In particular, temporal aggregation is a promising approach to intermittent demand forecasting, in which a lower-frequency time series can be aggregated to a higher-frequency time series. Latent characteristics of the demand, such as trend and seasonality, appear at higher levels of aggregation. \citet{nikolopoulos2011aggregate} first introduced temporal aggregation to intermittent demand forecasting and proposed the Aggregate-Disaggregate Intermittent Demand Approach (ADIDA). To tackle the challenge of determining the optimal aggregation level, \citet{petropoulos2015forecast} considered combinations of forecasts from multiple temporal aggregation levels simultaneously. This approach is called the Intermittent Multiple Aggregation Prediction Algorithm (IMAPA).  The overall results of their work suggested that combinations of forecasts from different frequencies led to improved forecasting performance.

Recently, some attention has been paid to applying machine learning approaches to improve forecasting accuracy for intermittent demand, such as neural networks \citep{lolli2017single}, support vector machines \citep{kaya2018intermittent, jiang2020intermittent}, and so on.

Despite that intermittent demand forecasting has obtained some research achievements in recent decades \citep{nikolopoulos2011aggregate,petropoulos2015forecast,kourentzes2021elucidate}, there is still much scope for improvements \citep{nikolopoulos2021we}. For example, limited attention has been given to combination schemes for intermittent demand forecasting.  The literature indicates that forecast combination can improve forecast accuracy in modeling fast-moving time series \citep{bates1969combination,de2000review, petropoulos2022forecasting,li2022improving}. In this study, we aim to examine whether the forecast combination improves intermittent demand forecasts.
The main contributions of our work are: (1) providing a discussion and comparison of forecast combination methods in the context of intermittent demand forecasting, (2) developing a feature-based combination framework for intermittent demand, which can determine optimal combination weights evaluated by the given error measure, and (3) improving the accuracy of both point and quantile forecasts to support real inventory decisions.

The rest of the paper is organized as follows. Section \ref{sec:combination} reviews a series of forecast combination methods discussed in this work. Section \ref{sec:DIVIDE} proposes a generalized forecast combination framework for intermittent demand.  In Section \ref{sec:Application}, we apply our framework to two real datasets and present results based on point forecasts and quantile forecasts. Section \ref{sec:conclusion} concludes the paper.

\section{A review of forecast combinations}
\label{sec:combination}

Combining forecasts from different methods or models has achieved satisfactory results in practice. \cite{wang2022forecast} provided an up-to-date review of forecast combinations including combining point forecasts and combining probabilistic forecasts. The Simple Average (SA) has been proved to be a hard-to-beat forecast combination method \citep{clemen1989combining, stock2004combination, lichtendahl2020some}, which simply combines forecasts with an equal weight of $1/M$ ($M$ is the number of forecasting methods to be combined). \citet{clemen1989combining} reviewed over two hundred articles and concluded that SA should be used as a benchmark when proposing more complex weighting schemes. \citet{palm1992combine} emphasized that SA could reduce the variance of forecasts and avoid the uncertainty of weight estimation. The phenomenon that SA outperforms more complicated combination methods is referred to ``forecast combination puzzle" \citep{stock2004combination, smith2009simple,claeskens2016forecast}.

Because SA is sensitive to extreme values, some attention has been paid to other more robust combination schemes, including the median and trimmed means \citep{stock2004combination, lichtendahl2020some, PETROPOULOS2020110}. \citet{jose2008simple} studied two mean-based methods, trimmed and Winsorized means, and verified their improved combined forecasts. The simple combination schemes based on the mean and median are easy to calculate and avoid parameter estimation error. However, there is still no consensus on which of the mean and the median of individual forecasts performs better.

In the field of intermittent demand forecasting, forecast combination methods have been largely overlooked. To the best of our knowledge, only SA has been applied to improving intermittent demand forecasting \citep{petropoulos2015forecast}. Recently, the organizers of the M5 competition \citep{makridakis2021m5} used SA as the combination benchmark, such as the average of exponential smoothing (ES) and ARIMA.  The  M5 competition focused on sales forecasts involving a mass of intermittent time series.
As shown in M5 results, combinations performed better or equally well with the individual methods that they consist of  \citep{makridakis2022m5}.

To further exploit the value of forecast combinations, a handful of research has focused on finding optimal weights for combining different forecasting models over the past half-century. The seminal work by \citet{bates1969combination} proposed the idea of weighted forecast combinations. \citet{newbold1974experience} continued this stream of research and investigated more forecasting models and multiple forecast horizons. In their work, a weighted combination can be expressed as a linear function such that
\begin{equation*}
\label{eq:yhat_c}
\hat{y}_{T+1}^{c}=\sum\limits_{i=1}^{M}{{{w}_{i,T+1}}{{{\hat{y}}}_{i,T+1}}}={\mathbf{w}_{T+1}'}{\mathbf{\hat{y}}_{T+1}},
\end{equation*}
where $\mathbf{\hat{y}}_{T+1}$ is the column vector of forecasts at time $T+1$ generated from $M$ forecasting models, and ${\mathbf{w}}$ is the column vector of weights.

\citet{granger1984improved} investigated some regressive approaches to obtain linear combinations. They demonstrated that the method with a constant term and unrestricted weights performed better. The combination weights can be estimated by Ordinary Least Squares (OLS). Linear combination has a long and successful history in forecasting. However, the issue related to determining the best set of forecasting models to combine is also worthy of attention. Lasso-based methods can do this trick by producing the selection and shrinkage toward zero \citep{tibshirani1996regression}. \citet{diebold2019machine} proposed a variant of Lasso, partially-egalitarian LASSO (peLASSO), which set the weights of some forecasting methods to zero and shrunk the survivors toward equality.  They provided an empirical assessment to forecast Eurozone GDP growth and found that peLASSO outperformed SA and the median \citep{diebold2019machine}.

The aforementioned weighted forecast combinations need to generate multiple forecasts in the training period, which multiplies the computation time. Especially for highly intermittent demand, the covariance matrix of forecast errors is often singular (can not calculate the inversion
 in \citet{bates1969combination}'s methods), because the obtained errors may have many zero values.
Similarly, for regressive approaches, the standardizing process can not be implemented when the true values for training are always zero.
Therefore, \citet{bates1969combination}'s weighted combination methods and regression-based methods are not applicable for highly intermittent data.

Recent studies indicate that using all time series in the dataset to estimate the combination weights show outstanding performance in forecasting fast-moving time series \citep[e.g.,][]{montero2021principles,talagala2022fformpp,wang2022uncertainty}. One mainstream is feature-based forecast combinations. For example, \citet{monteromanso2020fforma:} developed FFORMA (Feature-based FORecast Model Averaging), which used 42 features to estimate the optimal combination weights based on a meta-learning algorithm.

Different feature-based combination approaches applied different time series features to improve forecasting performance \citep[e.g.,][]{wang2009rule, petropoulos2014horses, kang2020gratis, li2022bayesian}. However, a significant characteristic of intermittent demand is that there exist a large number of zeros and irregular patterns, which makes the feature sets used in previous literature inapplicable for intermittent demand. \citet{theodorou2021exploring} proposed a methodological approach for feature extraction and selection to explore the representativeness of M5 dataset. On the basis of the FFORMA framework \citep{monteromanso2020fforma:}, \citet{kang2022forecast} used the diversity of forecasting models as the only feature. The diversity has proved to be a novel type of efficient feature, which can not only improve the forecast accuracy but also reduce the computational complexity.

The potential of time series features and the diversity of forecasts have not been investigated when producing forecast combinations for intermittent demand.  In our work, we extract a set of time series features selected for intermittent demand and calculate the diversity based on a pool of intermittent demand forecasting methods. To this end, a forecast combination framework for intermittent demand can be constructed by mapping the two types of features to the combination weights based on eXtreme Gradient Boosting (XGBoost) \citep{chen2016xgboost}. The proposed framework can be applied to both point and quantile forecast combinations.

\section{Forecast combination for intermittent demand}

\label{sec:DIVIDE}

\subsection{Time series features for intermittent demand}
\label{subsection: features}

Several studies have investigated the features of intermittent demand \citep{Kourentzes2016tsintermittent,Hara2021feasts,theodorou2021exploring}. First, we consider the two most popular attributes to divide intermittent demand in the SBC classification scheme \citep{syntetos2005categorization,kostenko2006note}. Then we review the 42 features selected for exploring the feature spaces of M5 competition data \citep{theodorou2021exploring}. To ensure the interpretability and compute as few features as possible,  we remove the features based on complex statistical methods, such as STL decomposition and Fourier transform, and take out Boolean variables with minimal information. The reserved features in  \cite{theodorou2021exploring} are used in our work.

Therefore, we consider nine explainable time series features for intermittent demand forecasting, which are listed in \autoref{tab: fide_features}.  They imply the intermittency, volatility, regularity and obsolescence of intermittent demand. Given a time series $\left\{ {{y}_{t}},t=1,2,\cdots ,T \right\}$, we describe the nine features as follows.

\begin{itemize}
\item ${F}_{1}$, ${F}_{2}$: The two features are average Inter-Demand Interval (\textsf{IDI}) and squared Coefficient of Variation (\textsf{CV}$^2$) to measure intermittency and demand size volatility in the SBC classification scheme \citep{syntetos2005categorization,kostenko2006note}.

\item ${F}_{3}$: Entropy-based measures have been applied to quantify the regularity and unpredictability of time-series data \citep{kang2017visualising, theodorou2021exploring}. We use the approximate entropy in this paper. A relatively small value of ${F}_{3}$ indicates that the demand series includes more regularity and is more forecastable.

\item ${F}_{4}$, ${F}_{5}$: The two features describe the ratios of some specific values in a given time series. ${F}_{4}$ measures the percentage of zero values. ${F}_{5}$ denotes the percentage of values lying outside $[\mu_y - \sigma_y, \mu_y + \sigma_y]$, where $\mu_y$ and $\sigma_y$ are the mean and standard deviation of time series $\{y_t\}$, respectively.

\item ${F}_{6}$: This feature provides the coefficient of a linear least squares regression, which measures the linear time trend of the variances of component chunks for the target series. For monthly data in the following experiments, we set the chunk length $L=12$. Moreover for daily data $L=10$, consistent with \cite{theodorou2021exploring}.

\item ${F}_{7}$: ${F}_{7}$ first calculates the consecutive changes of the demand, i.e., the first difference of the demand series. Then the mean absolute value of the consecutive changes is taken.

\end{itemize}

The last two features focus on the presence of recent demand to capture the obsolescence, which is a challenging problem in the field of intermittent demand \citep{babai2014intermittent,babai2019new}.

\begin{itemize}

\item ${F}_{8}$:  ${F}_{8}$ calculates the sum of squares of the last chunk out of $K$ chunks expressed as a ratio with the sum of squares over the whole series. We set $K = 4$ for the  Royal Air Force (RAF) dataset and $K = 10$ for M5 competition data, so that the length of the last chunk of each series is longer than the forecasting horizon.

\item ${F}_{9}$: ${F}_{9}$ computes the percentage of consecutive zero values at the end of the series, i.e., the number of consecutive zero values at the end over the length of the time series.
\end{itemize}

Based on the nine time series features tailored for intermittent demand, each target time series can be represented using a nine-dimensional vector. The feature vector can be used as the input to train the forecast combination model in the proposed framework.

\begin{table}
  \footnotesize
  \centering
  \caption{Description of nine time series features selected for intermittent demand.}
  \label{tab: fide_features}
  \begin{tabular}{lp{6cm}ll}
    \toprule
    Feature      & Description    & Range     & Implication   \\
    \midrule
    ${F}_{1}$: \textsf{IDI}
    & Averaged inter-demand interval
    & $\left[ \left. 1,\infty  \right) \right.$
    & Intermittency\\

    ${F}_{2}$: \textsf{CV$^2$}
    & Coefficient of variation squared of non-zero demand
    & $\left[ \left. 0,\infty  \right) \right.$
    & Volatility \\

    ${F}_{3}$: \textsf{Entropy}
     & Approximate entropy
     & $(0,1)$
     & Regularity\\

    ${F}_{4}$: \textsf{Percent.zero}
    & The percentage of observations that are zero values
    & $(0,1)$
    & Intermittency\\

    ${F}_{5}$: \textsf{Percent.beyond.sigma} & The percentage of observations that are more than $\sigma$ away from the mean of the time series ($\sigma$ is the standard
    deviation of the time series)
    & $\left[ \left. 0,1 \right) \right.$
    & Volatility\\

    ${F}_{6}$: \textsf{Linear.chunk.var}  &The coefficient of linear least-squares regression for variances of component chunks in the time series   & $\left( -\infty ,\infty  \right)$
    & Volatility\\

    ${F}_{7}$: \textsf{Change.mean.abs}
    & The mean absolute value of consecutive changes of the demand. & $\left[ \left. 0,\infty  \right) \right.$
    & Volatility \\

    ${F}_{8}$: \textsf{Ratio.last.chunk}  & The ratio of the sum of squares of the last chunk to the whole series
    & [0,1]
    & Obsolescence \\

    ${F}_{9}$: \textsf{Percent.zero.end}
    & The percentage of consecutive zero values at the end of the time series
    & [0,1]
    & Obsolescence\\
    \bottomrule
  \end{tabular}%
  \label{tab:ifeatures}%
\end{table}%

\subsection{Diversity for intermittent demand}
\label{subsection: diversity}

In this paper, we extend \citet{kang2022forecast}'s work and use the diversity of the pool of methods to develop a forecasting combination method for intermittent demand. The scaled diversity between any two forecasting methods is defined as:

\begin{equation}
\label{eq:Dij}
{{DIV}_{ij}}=\frac{\frac{1}{H}\sum\nolimits_{h=1}^{H}{{{\left( {{{\hat{y}}}_{ih}}-{{{\hat{y}}}_{jh}} \right)}^{2}}}}{{{\left( \frac{1}{T}\sum\nolimits_{t=1}^{T}{\left| {{y}_{t}} \right|}  \right)}^{2}}},
\end{equation}
where $H$ is the forecast horizon, ${{\hat{y}}}_{ih}$ is the $h$-th step forecast generated from the $i$-th forecasting model, and $\left\{ {{y}_{t}},t=1,2,\cdots ,T \right\}$ is a series of observed values.

Assuming that the forecasting pool contains $M$ methods, we apply \autoref{eq:Dij} to each two of them. For each target time series, we can construct a diversity vector consisting of $M(M - 1)/2$ pairwise diversity measures. This vector can be viewed as the feature vector for the corresponding series in the proposed framework.

The main merits of applying forecast diversity to intermittent demand forecasting are twofold.  The first aspect is simplicity in principle, as the calculation only depends on forecasting values, with no need to compute a separate set of features.  The second is general applicability. The diversity can be obtained automatically from intermittent demand forecasts and comprehended quickly by forecasters without expertise. Choosing relevant features to match the actual inventory management problem may lead to feature selection bias, especially when forecasters’ information is inadequate. Therefore, in contrast to time series features, the diversity shows remarkably simplicity and interpretability in intermittent demand forecasting.

\subsection{Evaluation metrics for intermittent forecasting}
\label{sec:error}

In previous studies, various forecasting evaluation metrics have been used for intermittent demand. The chosen metric of forecast errors may influence the ranked performance of the forecasting methods. \citet{silver1998inventory} pointed out that no single metric was universally best.  \citet{wallstrom2010evaluation} discussed a series of forecasting error measurements, especially for intermittent demand and split them into two categories, traditional (accuracy) and bias error measurements. As traditional measures, \citet{wallstrom2010evaluation} considered mean absolute deviation (MAD), mean square error (MSE), symmetric Mean Absolute Percentage Error (sMAPE). As bias error measures, they examined the Cumulated Forecast Error (CFE), Number Of Shortages (NOS), and Periods In Stock (PIS).
\cite{kourentzes2014intermittent} evaluated model selection results based
on two accuracy metrics. The first is the Mean Absolute Scaled Error (MASE), which
was suggested to be the standard measure for the data with different scales and zero values \citep{hyndman2006another}. The second is the scaled Absolute Periods In Stock (sAPIS), which is a scale-independent variant of PIS.

However, \cite{kolassa2016evaluating} explored traditional accuracy measures and argued that measures such as MAD, MASE and MAPE are unsuitable for intermittent demand. A flat zero forecast is frequently ``best'' for the measure of MAE when the demand is highly intermittent, because zero is the conditional median of the demand. Therefore, especially for intermittent demand, an MAE-minimizing method, which is the conditional median, prefers a lower forecast than the MSE-minimizing method, which is the expectation. In recent M5 competition with much intermittent demand, the accuracy of point forecasts is required to be evaluated using Root Mean Squared Scaled Error (RMSSE) \citep{makridakis2021m5}.

\cite{kolassa2020best} emphasized that different error measures reward different point forecasts, and different measures should not be applied to a single point forecast \citep{petropoulos2022forecasting}. In our work, we use RMSSE \citep{makridakis2021m5} to measure the performance of point forecasts, which can be obtained as:
 \begin{equation}
  \label{eq:RMSSE}
RMSSE=\sqrt{\frac{1}{H}\frac{\sum\nolimits_{h=1}^{H}{{{\left( {{y}_{T+h}}-{{{\hat{y}}}_{T+h}} \right)}^{2}}}}{\frac{1}{T-1}\sum\nolimits_{t=2}^{T}{{{\left( {{y}_{t}}-{{y}_{t-1}} \right)}^{2}}}}},
\end{equation}
where $H$ is the forecasting horizon. ${{\hat{y}}}_{T+h}$ is the $h$-th step forecast generated from a series of observed values $\left\{ {{y}_{t}},t=1,2,\cdots ,T \right\}$, and ${{y}_{T+h}}$ is the true value.

\subsection{Generalized forecast combination framework}
\label{framework}

The quality of forecast combination has been demonstrated to depend on the individual forecasts as well as the diversity between forecasts \citep{lemke2010meta, kourentzes2019another, kang2022forecast}. Therefore, defining an appropriate forecasting pool is one of the most crucial steps in the forecast combination process. Firstly, we define a broad pool for intermittent demand forecasting.
The pool includes traditional forecasting models, which are Naive, seasonal Naive (sNaive), Simple Exponential Smoothing (SES), Moving Averages (MA), AutoRegressive Integrated Moving Average (ARIMA), ExponenTial Smoothing (ETS), and intermittent demand forecasting methods, which are Croston’s method (CRO), optimized Croston’s method (optCro), SBA, TSB, ADIDA, IMAPA. The 12 forecasting methods in the pool are considered as statistical benchmarks in the M5 competition \citep{makridakis2021m5}. In contrast to CRO with fixed smoothing parameters, the parameters in optCro are optimized to allow for more flexibility.  Implementations for these methods exist in the \pkg{forecast} \citep{Hyndman2020forecast} and \pkg{tsintermittent} \citep{Kourentzes2016tsintermittent} packages in \proglang{R}.
Then the pooling methods \citep{kourentzes2019another, lichtendahl2020some, diebold2019machine} can be applied to reduce the number of forecasting methods and further improve the quality of the forecasting pool. We study the effect of three popular pooling algorithms in Section \ref{sec:Application}.

In the proposed forecast combination framework, we build an XGBoost model to learn the relationship between features and combination weights. This approach transforms the combination problem into a classification problem by setting the best forecasting method as the target class for each time series. The two types of features in Section \ref{subsection: features} and Section \ref{subsection: diversity} are all valid inputs for the forecast combination model.  We name the approach based on the nine time series features in Section \ref{subsection: features} as Feature-based Intermittent DEmand forecasting (FIDE).  The diversity-based method is called DIVersity-based Intermittent DEmand forecasting (DIVIDE).

Then, given a forecast error metric, the optimization objectives for the FIDE and DIVIDE are
\begin{subequations}
\label{eq:opt}
\begin{align}
&\underset{{w}_{F}}{\mathop{\arg \min }}\,\sum\limits_{n=1}^{N}{\sum\limits_{i=1}^{M}{w{{\left( {{F}_{n}} \right)}_{i}}}}\times \text{error}_{n,i}, ~ \textrm{and} \label{eqn:fide}\\
&\underset{{w}_{D}}{\mathop{\arg \min }}\,\sum\limits_{n=1}^{N}{\sum\limits_{i=1}^{M}{w{{\left( {{D}_{n}} \right)}_{i}}}}\times \text{error}_{n,i}, \label{eqn:divide}
\end{align}
\end{subequations}
respectively, where ${F}_{n}$ is the feature vector, ${D}_{n}$ is the diversity vector of the $n$-th time series, $N$ is the number of time series, and $M$ is the number of forecasting methods. $\text{error}_{n,i}$ is the forecast error of the $i$-th method for the $n$-th time series.  RMSSE is used as the error measure of point forecasts in this paper. RMSSE focuses on the expectation, which is consistent with the candidate methods in the forecasting pool. Once the model has been trained, weights can be produced for a new series to generate the combined forecast.
The process can be implemented based on the \proglang{R} package \pkg{M4metalearning} by \citet{monteromanso2020fforma:}.

\begin{figure}[h!]
	\centering
	\includegraphics[width=\textwidth]{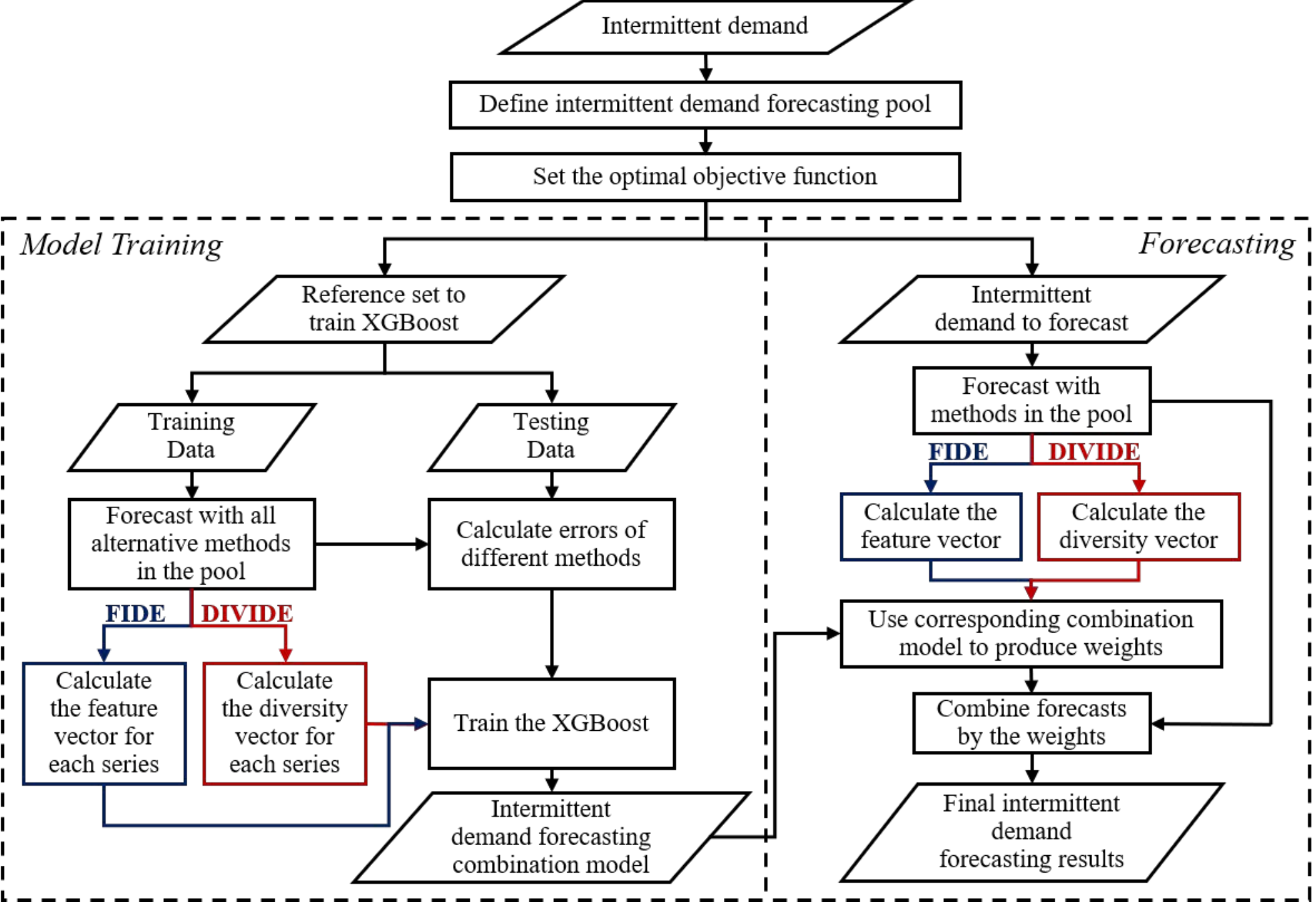}
	\caption
	{Caption: The flowchart of the proposed forecast combination framework for intermittent demand. \\
	Figure~\ref{fig:flowchart}. Alt Text: The flowchart contains two phases of model training and forecasting. The blue line refers to FIDE and the red line denotes DIVIDE. }
	\label{fig:flowchart}
\end{figure}

Based on the forecasting pool for intermittent demand, we put forward a generalized forecast combination framework containing FIDE and DIVIDE.  The flowchart of the proposed framework is presented in \autoref{fig:flowchart}.  In the training phase, we generate forecasts based on the methods in the intermittent demand forecasting pool and calculate errors required in the objective function. In FIDE, we compute the features selected for intermittent demand and learn the relationship between the features and combination weights by \autoref{eqn:fide}. In DIVIDE, the combination model can be obtained based on the diversity of different forecast methods (see \autoref{eqn:divide}), where the pairwise diversity values of the methods in the pool are used as time series features. Therefore, DIVIDE can be viewed as a special case of FIDE. In the forecasting phase, we calculate the features or the diversity for the new time series, and get the combination weights through the pre-trained XGBoost model. Finally, we utilize the optimal weights to average the forecasts from different methods in the pool and achieve the combined forecast results.

To evaluate the forecasting performance of the proposed framework, the time series need to be divided into three periods. Let $H$
be the forecast horizon and $T$
be the length of data. The first $T-H$ observations are used for training the forecast combination model. Then the $T-H$ observations are split into $T-H-H$ for training the forecasting methods in the pool and $H$ for testing. The final $H$ observations are used to evaluate the forecasting results.

The merits of the proposed framework include: (1) using a diverse forecasting pool, consisting of intermittent demand forecasting methods and traditional time series forecasting models, (2) considering a customizable objective function depending on actual inventory management requirements, (3) selecting intelligible time series features especially for intermittent demand, and (4) calculating the diversity with simple form only based on forecasting values.

\section{Empirical evaluation}
\label{sec:Application}

\subsection{Real dataset}

The proposed methods are applied to two real datasets. The first RAF dataset has been previously investigated in the literature \citep{kourentzes2021elucidate,petropoulos2015forecast,teunter2009forecasting}.  It contains 5000 monthly time series, with 84 observations each. Moreover, the second is M5 competition data, involving the unit sales of 3049 products sold by Walmart in the USA  between 2011-01-29 and 2016-06-19 (1969 days). The dataset was organized in the form of hierarchical time series in M5 competition \citep{makridakis2021m5}. We only consider the bottom level, i.e., 30,490 product-store unit sales in this paper.

In the following experiment, we examine three forecast horizons of 3, 6 and 12 months ahead for RAF dataset and 28-day-ahead forecasts for M5 data as required in M5 competition \citep{makridakis2021m5}. The final observations of the horizon length are used to evaluate the forecasting performance.
The seasonal periods are 12 and 7 for monthly and daily demand, respectively. Moreover, we preprocess the data before forecasting, removing the initial zero values and making the first non-zero demand as the initial value \citep{theodorou2021exploring}.
This is due to a lack of information that the initial zeros mean demands or sales are zero, or the product has not been in stores yet.

\begin{figure}
	\begin{center}
		\begin{tabular}{c}
			\includegraphics[width=\textwidth]{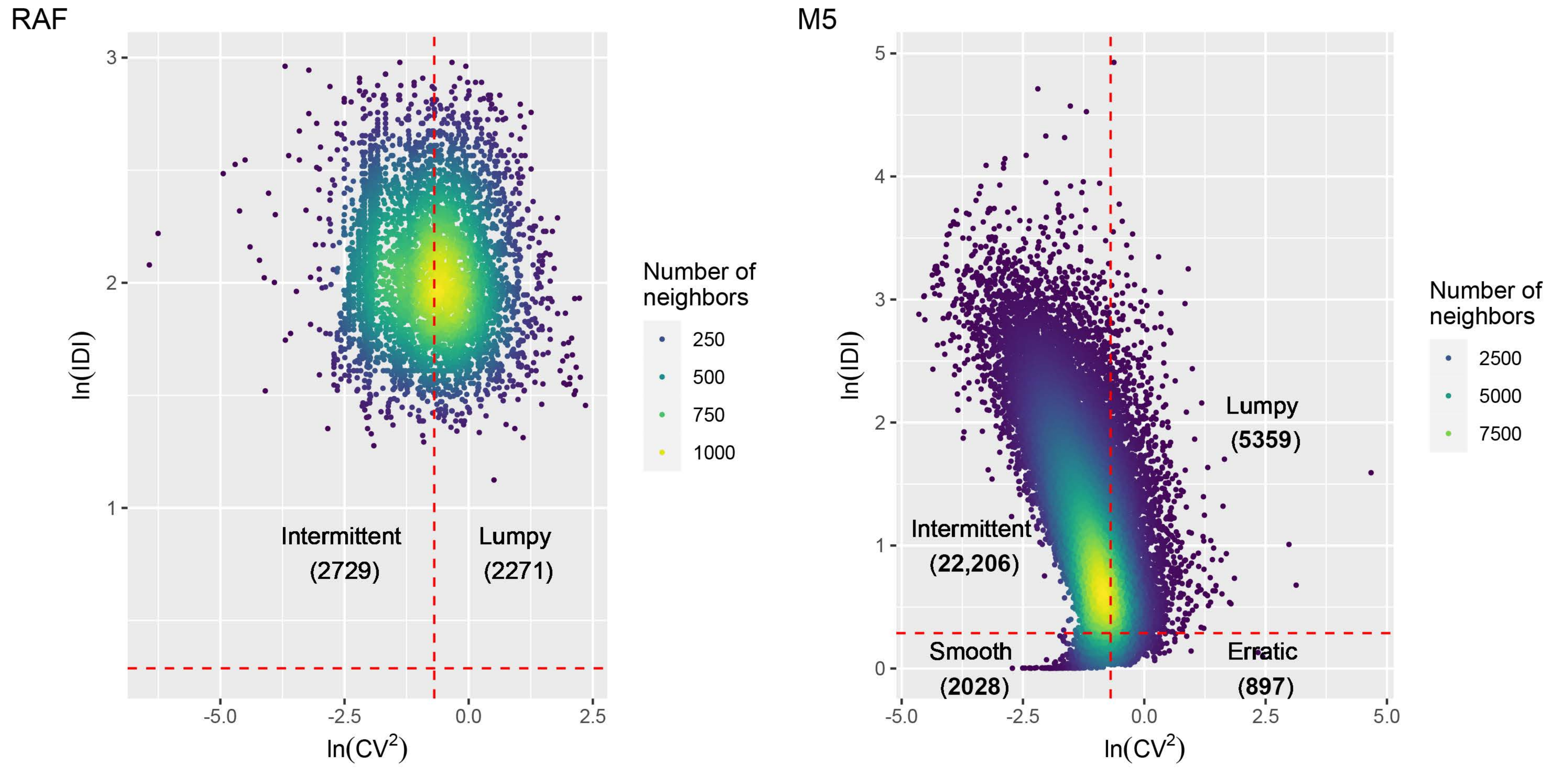}
		\end{tabular}
	\end{center}
	\caption
	{Caption: The density scatterplots of $\mathrm{ln(\textsf{CV}^2)}$ and $\mathrm{ln(\textsf{IDI})}$, for RAF (left) and M5 (right) datasets.  \\
	Figure~\ref{fig:data}. Alt Text: The x-axis and y-axis show the natural logarithmic transform of $\textsf{CV}^2$ and that of \textsf{IDI}, respectively, for RAF (left) and M5 (right) datasets. To present the overall distribution of data, each point is colored by the number of neighboring points. Red lines indicate the boundaries of different categories ($\textsf{IDI} = 4/3$, $\textsf{CV}^2 = 0.5$) \citep{kostenko2006note}. In the left panel, the RAF dataset contains 2729 intermittent, and 2271 lumpy series. In the right panel, the M5 data includes  22,206 intermittent, 5359 lumpy, 897 erratic, and 2028 smooth series. }
	\label{fig:data}
\end{figure}

Following the SBC scheme \citep{kostenko2006note}, the time series can be divided into four categories based on \textsf{IDI} and \textsf{CV}$^2$. \autoref{fig:data} describes the distributions of RAF and M5 datasets, respectively. The boundaries of different categories in \autoref{fig:data} are \textsf{IDI}$ = 4/3$, \textsf{CV}$^2 = 0.5$ \citep{kostenko2006note}.
The equal sign is placed on the less-than sign when classifying, e.g., if \textsf{IDI}$ > 4/3$ and \textsf{CV}$^2 <= 0.5$, the time series is ``intermittent''.
As shown in \autoref{fig:data}, the RAF dataset exhibits high intermittence and contains 2729 intermittent, and 2271 lumpy series. While the M5 data has a wider-ranging distribution, including 22,206 intermittent, 5359 lumpy, 897 erratic, and 2028 smooth series.

\subsection{Point forecasting}

We compare our methods with individual models, SA, Median and FFORMA.
Other combination methods reviewed in Section \ref{sec:combination}, such as \cite{bates1969combination}'s original forecast combinations and regression-based methods \citep{granger1984improved,diebold2019machine}, are omitted here, which are not applicable for highly intermittent data. We present the forecasting accuracy of different methods based RAF dataset and M5 competition data in \autoref{tab:pointforecast_raf} and \autoref{tab:pointforecast_M5} respectively.

As shown in \autoref{tab:pointforecast_raf},  the best individual method is ADIDA for all forecast horizons based on RMSSE. The simple combination methods (SA and Median) can not beat the best individual method. While the proposed methods based on intermittent demand features and the diversity consistently outperform others. FIDE shows obvious superiority compared with FFORMA using 42 time series features and offers the best forecasting results overall. Therefore, our chosen features are more appropriate to describe intermittent demand compared with the time series features in FFORMA designed for fast-moving data.  Furthermore, the improved forecast accuracy of DIVIDE indicates that the diversity is a simple and efficient tool for intermittent demand forecasting combinations.

The forecasting results of M5 competition data in \autoref{tab:pointforecast_M5} are organized by SBC classification scheme \citep{kostenko2006note}.
For each column (a subset of data), the combination model is optimized respectively. The last three rows show the RMSSE of the top three winning methods in the M5 competition for comparison, e.g., ``M5-w1'' denotes the first ranked method. The results of last three rows in \autoref{tab:pointforecast_M5} are calculated based on  corresponding M5 submissions.
It should be noted that the M5 competition took the hierarchy of  data into consideration and used weighted RMSSE to rank participants. The weights put more emphasis on the series that account for higher monetary sales \citep{makridakis2021m5}. Therefore, comparing these methods in absolute terms is not entirely fair, as they were obtained from different application contexts and optimization objectives.

As shown in \autoref{tab:pointforecast_M5}, the best individual method is IMAPA for all classifications, which is inconsistent with the RAF data. The results emphasize the risk of choosing a single forecasting method and elicit the necessity of forecast combination. Our proposed methods achieve the competitive forecasting performance based on RMSSE when compared with the top three ranked methods in the M5 competition. Based on different classifications of M5 data, the performance of the proposed methods exhibits significant differences. The proposed FIDE and DIVIDE  outperform FFORMA for the intermittent and lumpy data, which is consistent with the RAF dataset. The forecasting results based on the two datasets provide good evidence for the superiority of the proposed framework in intermittent demand forecasting. However, for the erratic and smooth data, our methods perform slightly worse than FFORMA. We acknowledge the limitations of the features used in the proposed framework, which are more applicable to intermittent demand.

\begin{table}
	\footnotesize
	\newcommand{\tabincell}[2]{\begin{tabular}{@{}#1@{}}#2\end{tabular}}
	\centering
	\caption{Forecasting accuracy (RMSSE) of different methods based on the RAF dataset. For each forecasting horizon (column), the smallest value is marked in \textbf{bold}. The last column is the average rank of all methods based on the three horizons.}
	\begin{tabular}{l ccccc }
		\toprule
		Method & $H=3$  &  $H=6$  &  $H=12$   & Average rank \\

		\midrule
		Naive         & 0.493 & 0.552 & 0.658 & 13.0 \\
		sNaive        & 0.619 & 0.764 & 0.911 & 17.0 \\
		SES           & 0.466 & 0.540 & 0.641 & 8.3  \\
		MA            & 0.461 & 0.551 & 0.644 & 9.0  \\
		ARIMA         & 0.490 & 0.558 & 0.616 & 12.0 \\
		ETS           & 0.483 & 0.552 & 0.615&  9.2  \\

		CRO           & 0.500 & 0.564 & 0.616 & 14.0 \\
		optCro        & 0.500 & 0.564 & 0.616 & 14.0 \\
		SBA           & 0.499 & 0.562 & 0.615 & 12.2 \\
		TSB           & 0.487 & 0.554 & 0.612 & 10.0 \\
		ADIDA         & 0.464 & 0.539 & 0.606 & 5.0  \\
		IMAPA         & 0.478 & 0.545 & 0.603 & 5.8  \\

		\midrule
		SA            & 0.485 & 0.552 & 0.618 & 11.0 \\
		Median        & 0.478 & 0.546 & 0.605 &	6.5  \\
		FFORMA        & 0.413 & 0.491 & 0.583 & 3.0 \\
		\textbf{FIDE} & 0.369  & \textbf{0.461} & \textbf{0.562} &   \textbf{1.3}      \\

		\textbf{DIVIDE} & \textbf{0.359} & 0.462 & 0.563& 1.7\\

		\bottomrule
	\end{tabular}%
	\label{tab:pointforecast_raf}%
\end{table}%

\begin{table}
	\footnotesize
	\newcommand{\tabincell}[2]{\begin{tabular}{@{}#1@{}}#2\end{tabular}}
	\centering
	\caption{Forecasting accuracy (RMSSE) of different methods based on the M5 competition dataset. The last three rows show the RMSSE of the top three winning methods in the M5 competition for comparison.
	The results for different classifications and the whole data set are presented respectively.
	For each column, the smallest value is marked in \textbf{bold} (without including the last three rows).}
	\begin{tabular}{l cccccc }
		\toprule
		Method & Intermittent  &  Lumpy  &  Erratic   & Smooth & All \\

		\midrule
		Naive  & 0.888 & 0.835 & 0.917 & 1.001 & 0.887\\
		sNaive & 0.892 & 0.840 & 0.929 & 1.024 & 0.892\\
		SES    & 0.835 & 0.794 & 0.870 & 0.925 & 0.835\\
		MA     & 0.859 & 0.807 & 0.884 & 0.960 & 0.857\\
		ARIMA  & 0.820 & 0.782 & 0.842 & 0.898 & 0.819\\
		ETS    & 0.822 & 0.790 & 0.868 & 0.920 & 0.824\\

		CRO    & 0.814 & 0.782 & 0.860 & 0.930 & 0.817\\
		optCro & 0.810 & 0.782 & 0.845 & 0.850 & 0.809\\
		SBA    & 0.811 & 0.784 & 0.841 & 0.852 & 0.810\\
		TSB    & 0.809 & 0.780 & 0.838 & 0.848 & 0.807\\
		ADIDA  & 0.823 & 0.790 & 0.870 & 0.925 & 0.826\\
		IMAPA  & 0.806 & 0.775 & 0.832 & 0.842 & 0.804\\

		\midrule
		SA     & 0.813 & 0.773 & 0.816 & 0.865 & 0.809  \\
		Median & 0.819 & 0.784 & 0.850 & 0.901 & 0.819	\\
		FFORMA & 0.809 & 0.747 & \textbf{0.739} & \textbf{0.775} &0.801 \\
		\textbf{FIDE}   & 0.792  &  0.745
		&  0.746 &    0.783   & 0.783\\
		\textbf{DIVIDE} & \textbf{0.790} & \textbf{0.741} & 0.743 & 0.776 & \textbf{0.779}\\

		\midrule
		M5-w1 & 0.785 & 0.734 & 0.743 & 0.774 & 0.774\\
		M5-w2 & 0.799 & 0.747 & 0.767 & 0.797 & 0.789\\
		M5-w3 & 0.784 & 0.731 & 0.730 & 0.769 & 0.772\\
		\bottomrule
	\end{tabular}%
	\label{tab:pointforecast_M5}%
\end{table}%

\begin{figure}
	\begin{center}
		\begin{tabular}{c}
			\includegraphics[width=\textwidth]{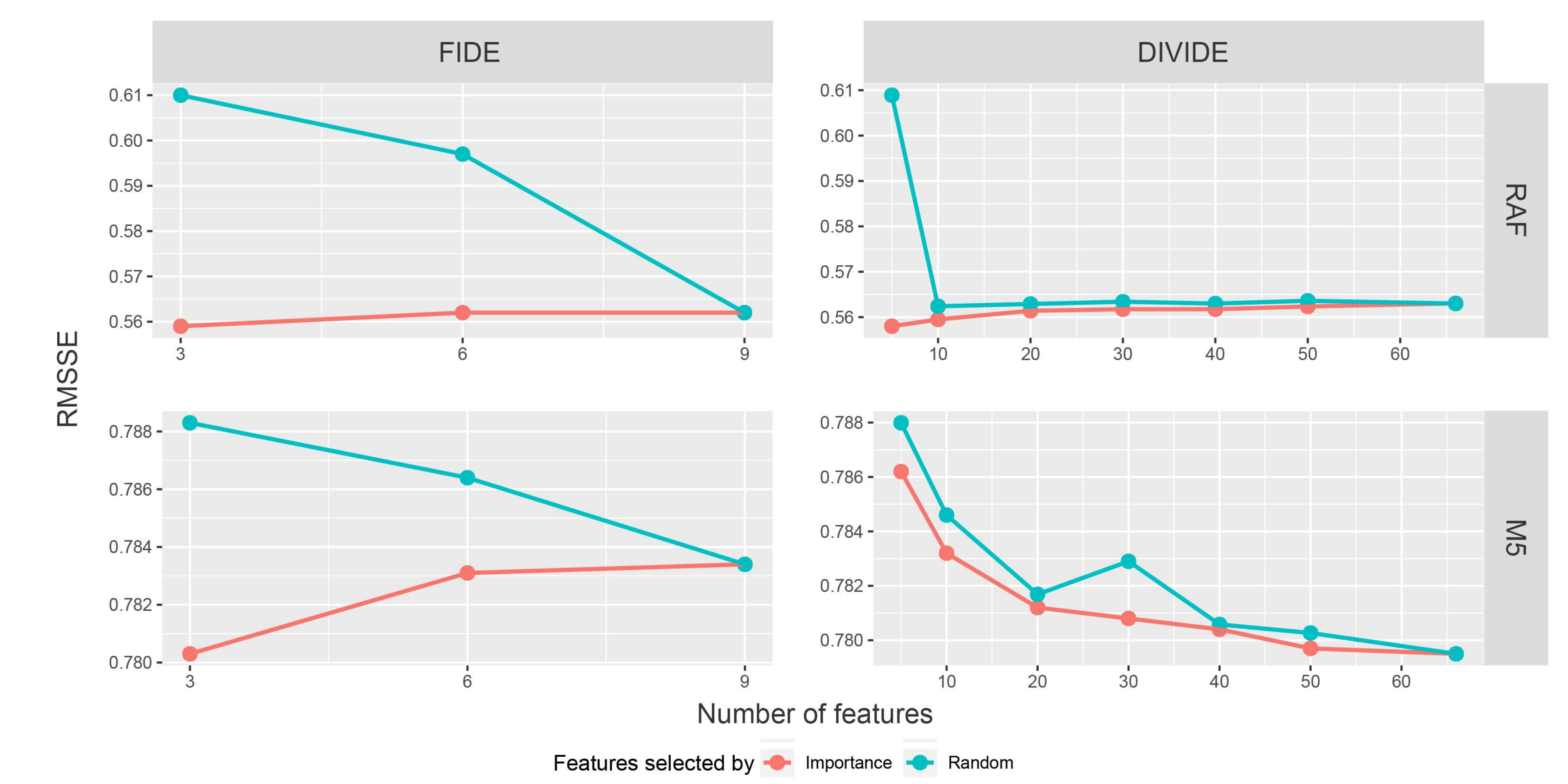}
		\end{tabular}
	\end{center}
	\caption
	{Caption: The relationship between RMSSE and the number of features used in the proposed FIDE (left) and DIVIDE (right), for RAF (top) and M5 (bottom) datasets.  \\
	Figure~\ref{fig:feature}. Alt Text: The results based on RAF dataset (indicatively for $H=12$) are presented on the top panel, and the bottom panel shows the results from M5 competition data. The features in FIDE and DIVIDE are selected based on two methods, one is to select features in order of the importance in XGBoost model (red lines), and the other is by random selection (blue lines).
	}
	\label{fig:feature}
\end{figure}

The features and diversity in the proposed framework have been proven to improve the accuracy of intermittent demand forecasting. In the following experiments, we continue to provide a sensitivity analysis of  multiple features, including diversity viewed as another form of features.
We investigate the relationship between RMSSE and the number of features used in the proposed FIDE and DIVIDE, as shown in \autoref{fig:feature}. In FIDE, we set the feature number to be 3, 6 and 9(all). While in DIVIDE, the feature number varies across 5, 10, 20, 30, 40, 50 and 66(all). We use two ways to alter the number of features. One is to select features in the order of feature importance in XGBoost model, and the other is by random feature selection. The importance of each feature to the FIDE or DIVIDE is measured by the gain of features in the XGBoost model \citep{chen2016xgboost}.

As shown in \autoref{fig:feature},  with the increase of the feature number, RMSSE of FIDE decreases when selecting features randomly (blue lines), but increases when selecting features in order of the importance (red lines). The findings emphasize the importance of choosing appropriate features as input in the proposed FIDE. For RAF dataset, the three most important features are \textsf{Percent.zero.end} ($F_9$), \textsf{Ratio.last.chunk} ($F_8$) and  \textsf{Linear.chunk.var} ($F_6$). While for M5 competition data, the top three features are \textsf{Percent.zero.end} ($F_9$), \textsf{IDI} ($F_1$) and \textsf{Ratio.last.chunk} ($F_8$). Thus, the features to capture the recent demand are more critical for constructing the combination model in FIDE.
However, the relationship between RMSSE and the feature number in DIVIDE is markedly different from that of FIDE. As the number of features increases, the overall trend of RMSSE is downward, though there is a non-significant increase when considering the importance of features for RAF dataset. Therefore, we recommend applying the whole diversity in the proposed framework.

\begin{figure}
	\begin{center}
		\begin{tabular}{c}
			\includegraphics[width=\textwidth]{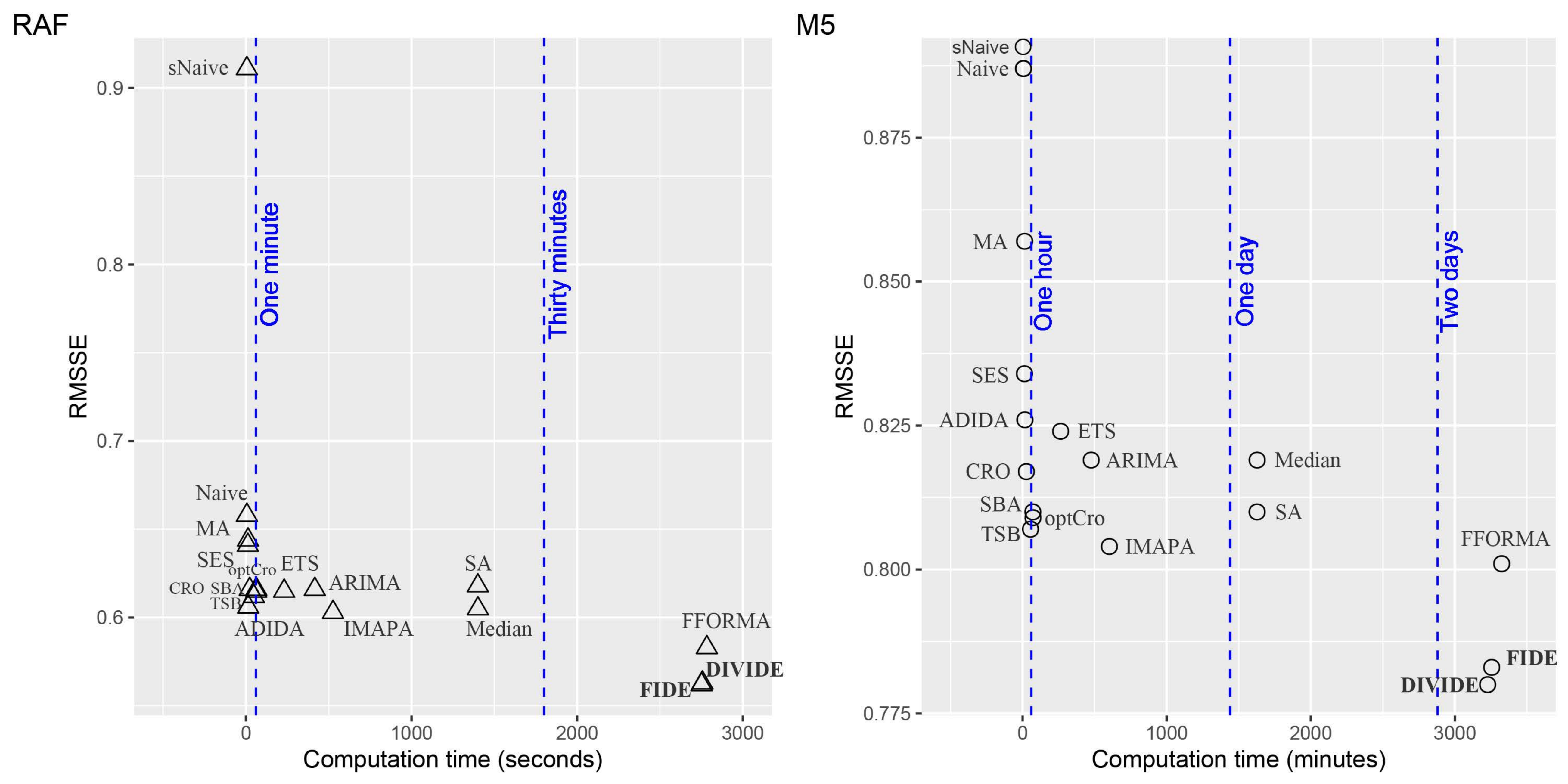}
		\end{tabular}
	\end{center}
	\caption
	{Caption: The relationship between RMSSE and computation time, for RAF (left) and M5 (right) datasets.  \\
	Figure~\ref{fig:time}. Alt Text: Different forecasting methods are marked on the picture. The computation time is obtained by using a Microsoft Windows 10 desktop with 8 cores and 16 logical processors at 3.59 GHz, 16 GB RAM. }
	\label{fig:time}
\end{figure}

To analyze the efficiency of the examined forecasting methods, we investigate the relationship between RMSSE and computation time based on RAF and M5 datasets.  The results are computed indicatively for RAF dataset when $H=12$.
As shown in \autoref{fig:time}, the time consumption of our methods mainly comes from the individual forecasting methods, which is the limitation of forecast combinations. Simple combination schemes, such as SA and median, can save nearly half the time, but they perform worse than the best individual method for intermittent demand. The proposed framework generates forecasts in the training and forecasting periods, increasing the time consumption. However, the process of model training is in the off-line phase in real applications. Therefore, the increased computational time of training does not affect the forecasting efficiency. Compared with FFORMA, our methods based on features for intermittent demand and diversity are more computationally efficient, especially for the M5 dataset with large amounts of long time series.
Based on these findings,  decision makers can consider the trade-off between accuracy and computational cost in actual inventory management.

Reducing the number of forecasting methods used in the proposed framework can significantly save computational time. For instance, the computation time of our methods can be halved by removing the three most time-consuming methods in the forecasting pool. In the following experiment, we aim to investigate the potential of shrinking the forecasting pool to further improve the accuracy of the proposed framework. We call this process pooling, deriving from \cite{kourentzes2019another}'s research.

We study the effect of three pooling algorithms based on the RAF (indicatively for $h=12$) and M5 dataset, which are forecast islands proposed by \citet{kourentzes2019another}, a screened method from \citet{lichtendahl2020some}, and a Lasso-based method by \citet{diebold2019machine}.  The forecast islands \citep{kourentzes2019another} remove some poorly performing models from the pool, which is shortened to ``Islands". It conducts ${C}'=\left\{ 0,\Delta C \right\}$ for a series of ordered forecasts based on a criterion of forecasting performance $C$ and includes all forecasts until ${C}'\ge T$. $T=\text{Q3}+1.5\text{IQR}$ is related to the outlier of the boxplot, where Q3 is the third quartile and IQR is the interquartile range. The screened method \citep{lichtendahl2020some}, shortened to ``Screened", screens out forecasting models with highly correlated errors (correlation coefficient is over 0.95). The Lasso-based method \citep{diebold2019machine}, ``Lasso'' for short, sets the regression coefficients of some forecasts to zero via a standard Lasso software, i.e., \proglang{R} package \pkg{glmnet} \citep{Friedman2010Regularization}, and the survivors form the final forecast pool.

\autoref{tab:pool} presents the forecasting accuracy of original FIDE and DIVIDE and those considering the three pooling algorithms.  In \autoref{tab:pool}, we find that none of the pooling methods significantly improves the forecasting performance. The proposed framework automatically reduces the weights of some methods to minimal values, which can be regarded as a generalized pooling method customized for each time series. The Islands remove the worst performing methods in the pool, but reduce the accuracy, especially for RAF dataset. For highly intermittent data, the poorly performing methods, such as Naive, sNaive, make important contributions to the forecast combination.
Therefore, unless the pool of forecasting methods is too large which would render the computation process time-consuming, there is no need to add modeling complexity by implementing pooling approaches on top of our proposed framework.

\begin{table}
	\footnotesize
	\newcommand{\tabincell}[2]{\begin{tabular}{@{}#1@{}}#2\end{tabular}}
	\centering
	\caption{Forecasting accuracy (RMSSE) of the original FIDE and DIVIDE methods and those considering pooling algorithms based on RAF and M5 dataset.  For each row (dataset), the smallest values for FIDE and DIVIDE are marked in \textbf{bold}, respectively.
	}
	\begin{tabular}{l cccc cccc}
		\toprule
		& \multicolumn{4}{c}{FIDE} & \multicolumn{4}{c}{DIVIDE}    \\
		\cmidrule(lr){2-5} \cmidrule(lr){6-9}
		& Original   & Islands & Screened & Lasso
		& Original   & Islands & Screened & Lasso \\
		\midrule
		RAF($H=12$) & 0.562 & 0.600 & 0.562 & \textbf{0.561}
		    & 0.563 & 0.597 & 0.561 & \textbf{0.560}\\
		M5 & \textbf{0.783} & 0.787 & \textbf{0.783} & 0.784
		   & \textbf{0.779} & 0.783 & 0.781 & 0.782\\
		\bottomrule
	\end{tabular}%
	\label{tab:pool}%
\end{table}%

\subsection{Quantile forecasting}
\label{Quantile forecast}

Based on the improving accuracy of point forecasts, the proposed combination methods are shown to be effective in providing robust forecasts to support decisions. However, in real supply chain management,  estimating the right part of the demand distribution is also necessary for determining safety stock levels, which has been largely ignored in the research \citep{barrow2016distributions, spiliotis2021product}.  \cite{fildes2019retail} reviewed retail demand forecasting and emphasized the connection of quantile, density, or volatility forecasting to the inventory control.

The intermittent demand forecasting methods (such as CRO, optCro, SBA, TSB, ADIDA and IMAPA) can not directly output quantile forecasts. We apply \cite{trapero2019empirical}'s empirical approach to estimate the desired quantiles. They recommended a kernel density estimation to model the forecast error distribution. We generate quantile forecasts by adjusting the point forecasts  based on the respective quantiles calculated from the empirical distribution of residual errors, as follows:
\begin{equation}
  \label{eq:quantile}
{{Q}_{T+h}}\left( u \right) = {\hat{y}}_{T+h} + {\hat{q}_{|e}}\left( u \right),
 \end{equation}
where $T$ is the length of observations, ${{Q}_{T+h}}\left( u \right)$ is the probabilistic
forecast for quantile $u$ at time $T+h$, ${\hat{y}}_{T+h}$ is the $h$-th step point forecast, ${\hat{q}_{|e}}\left( u \right)$ is the estimated $u$-th quantile of the residual errors. As shown in \autoref{eq:quantile}, we assume that the demand pattern that occurred in the past will continue in
the future.  The obtained forecasts are based on in-sample approximations without requiring computing multiple forecasts. The approach has been verified to perform well for the RAF and M5 datasets
 \citep{spiliotis2021product, kourentzes2021elucidate}.

The proposed framework can be extended to quantile forecast combinations by mapping the features to the errors of quantile forecasts.  In FIDE and DIVIDE, we still compute the nine features in Section \ref{subsection: features} based on historical data and the diversity of different point forecasts as shown in Section \ref{subsection: diversity}. The training and testing processes are consistent with Section \ref{framework}. We use the Scaled Pinball Loss (SPL) function to measure the precision of the quantile forecasts, which is required in  the M5 competition \citep{makridakis2021m5}. The SPL can be obtained as follows:

\begin{footnotesize}
 \begin{equation}
  \label{eq:SPL}
  \begin{aligned}
& SPL(u)= \\
& \frac{\sum\limits_{h=1}^{H}{u\left( {{y}_{T+h}}-{{Q}_{T+h}}\left( u \right) \right)\mathbf{1}\left\{ {{Q}_{T+h}}\left( u \right)\le {{y}_{T+h}} \right\}+\left( 1-u \right)\left( {{Q}_{T+h}}\left( u \right)-{{y}_{T+h}} \right)\mathbf{1}\left\{ {{Q}_{T+h}}\left( u \right)>{{y}_{T+h}} \right\}}}{H\cdot\frac{1}{T-1}\sum\nolimits_{t=2}^{T}{\left| {{y}_{t}}-{{y}_{t-1}} \right|}}, \\
 \end{aligned}
 \end{equation}
 \end{footnotesize}
where ${{y}_{T+h}}$ is the actual future value of the examined time series at point $T+h$, ${{Q}_{T+h}}\left( u \right)$ is the generated forecast for quantile $u$, $H$ is the forecasting horizon, $T$ is the length of the number of historical observations, and \textbf{1} is the indicator function (being 1 if true is within the postulated interval and 0 otherwise).

The following experiment based on the RAF and M5 datasets focuses on
four quantiles, i.e. $u_1 = 0.750$, $u_2 = 0.835$, $u_3 = 0.975$, and $u_4 = 0.995$.  $u_1$ and $u_2$ provide a good sense of the mid-right part of the distribution, while $u_3$ and $u_4$ provide information about its right tail, which is essential for the risk of extreme outcomes.  We customize the objective function by assigning the  error measure to SPL based on the correlated quantiles. Therefore, we obtain different combination weights for the four quantiles, respectively. The forecasting results in \autoref{tab:quantile} are computed based on 12-month-ahead forecasts for RAF dataset and 28-day-ahead forecasts for M5 data.

We can find in \autoref{tab:quantile} that the performance of individual methods changes considerably based on different quantiles. The finding indicates that each method is more appropriate for estimating different parts of the distribution of the series,  which echoes the weakness of choosing a single method. The proposed combination methods exhibit steady performance across the four quantiles. For RAF dataset, FIDE and DIVIDE consistently outperform the rest in \autoref{tab:quantile}. For M5 dataset, we add the top three ranked methods in the M5 competition for comparison, the results of which derive from \cite{spiliotis2021product}. The proposed DIVIDE outperforms the first ranked method in M5 competition at quantiles 0.835, 0.975 and 0.995. While at quantile 0.750, our methods also provide competitive forecasting results. The improved performance of high quantile forecasts can contribute to practical inventory decisions for higher levels of service.

\begin{table}
  \footnotesize
  \newcommand{\tabincell}[2]{\begin{tabular}{@{}#1@{}}#2\end{tabular}}
  \centering
  \caption{Quantile forecasting performance (SPL) of different methods based on the RAF and M5 datasets.  The last three rows show the SPL of the top three winning methods in the M5 competition for comparison.
  The results based on four quantiles (0.750, 0.835, 0.975, and 0.995) are reported. For each column (quantile), the smallest value is marked in \textbf{bold} (without including the last three rows).}
  \begin{tabular}{l cccc cccc  }
    \toprule
     \multicolumn{1}{l}{\multirow{2}{*}{Method}} & \multicolumn{4}{c}{RAF} & \multicolumn{4}{c}{M5} \\
    \cline{2-5}
    \cline{6-9}
    & $0.750$ & $0.835$ & $0.975$  & $0.995$
   & $0.750$ & $0.835$ & $0.975$  & $0.995$\\

    \midrule
    Naive  & 1.395 & 1.296 & 0.455 &  0.183
           & 1.078 & 1.035 & 0.329 & 0.090\\
    sNaive & 0.844 & 0.778 & 0.353 &  0.211
           & 0.636 & 0.557 & 0.181 & 0.063\\
    SES    & 0.864 & 0.793 & 0.353 &  0.207
           & 0.576 & 0.510 & 0.214 & 0.110\\
    MA     & 0.503 & 0.509 & 0.351 &  0.205
           & 0.599 & 0.561 & 0.204 & 0.068\\
    ARIMA  &  0.740 & 0.685 & 0.354 &  0.237
           & 0.586 & 0.514 & 0.213 & 0.109\\
    ETS    & 0.741  & 0.687 & 0.355 &  0.238
           & 0.579 & 0.509 & 0.215 & 0.113\\
    CRO    & 0.404 & 0.447 & 0.348 &   0.192
           & 0.580 & 0.519 & 0.171 & 0.053\\
    optCro & 0.406 & 0.448 & 0.349 &   0.192
           & 0.566 & 0.510 & 0.169 & 0.053\\
    SBA    & 0.405 & 0.448 & 0.349 &   0.192
           & 0.566 & 0.509 & 0.169 & 0.053\\
    TSB    & 0.406 & 0.449 & 0.348 &   0.192
           & 0.566 & 0.510 & 0.169 & 0.053\\
    ADIDA  & 0.583 & 0.560 & 0.405 &   0.343
           & 0.568 & 0.520 & 0.292 & 0.199\\
    IMAPA  & 0.404 & 0.455 & 0.367 &   0.217
           & 0.561 & 0.504 & 0.170 & 0.056\\

    \midrule
    SA     & 0.629 & 0.609 & 0.347 &   0.198
           & 0.569 & 0.500 & 0.168 & 0.057\\
    Median & 0.532 & 0.531 & 0.345 &   0.199
           & 0.556 & 0.496 & 0.171 & 0.056\\
    FFORMA & 0.401 & 0.450 & 0.343 &   0.188
           & 0.523 & 0.471 & 0.156 &  0.051\\
    FIDE   & 0.402 & \textbf{0.446} & \textbf{0.340} &   \textbf{0.182}
           &  0.516 & 0.456 & 0.150 & 0.046\\
    DIVIDE & \textbf{0.400} & 0.449 & 0.341 &  0.184
           & \textbf{0.511} &  \textbf{0.453} & \textbf{0.146} & \textbf{0.045}\\

    \midrule
    M5-w1 & & & &
          & 0.509 & 0.455 & 0.151 & 0.048\\
	M5-w2 & & & &
	      & 0.610 & 0.492 & 0.157 & 0.055\\
	M5-w3 & & & &
	      & 0.513 & 0.457 & 0.165 & 0.070\\
    \bottomrule
  \end{tabular}%
  \label{tab:quantile}%
\end{table}%

\section{Conclusion}
\label{sec:conclusion}

This paper focuses on forecast combinations for intermittent demand. We review a handful of forecasting methods, and investigate the performance of some existing forecast combination methods for intermittent demand.  We introduce time series features and diversity to propose a generalized forecast combination framework, which can automatically determine the optimal combination weights.  We conduct an empirical investigation based on real-life data to analyze the forecast accuracy and gain insights related to inventory decisions.

The results of point forecasts are measured by RMSSE, which focuses on the expectation. The proposed framework notably outperforms other combination methods and the best individual method, especially for the RAF dataset with highly intermittent series. Moreover, for M5 competition data, our methods achieve a competitive performance compared with the top three ranked methods in the M5 competition. In addition, the proposed framework can be regarded as a generalized pooling method customized for each time series by reducing the weights of some methods to minimal values. The empirical evaluation based on RAF and M5 datasets provides good evidence of the superiority and flexibility of the proposed framework.
We acknowledge that our combination methods increase the computational time compared with individual methods.  Decision makers should consider the trade-off between accuracy and computational cost in actual inventory management.

The proposed framework has also been applied to quantile forecast combinations, especially for high quantiles to estimate the right part of the demand distribution. We use SPL to measure the quantile forecasting performance and make it used in the optimization objective. The examined results show that our methods can provide accurate forecasts of both central tendency and high quantiles, which directly connect with the inventory decision.

The good performance of our proposed framework can be attributed to: (i) defining a forecasting pool suitable for intermittent demand, which consists of several intermittent demand forecasting methods and traditional time series forecasting models, (ii) applying diversity and time series features to determine the optimal combination weights automatically, and (iii) applying to both point and quantile forecasts to support inventory decisions. The diversity and the features selected for intermittent demand are all effective inputs of the proposed framework. Extracting the diversity independent of historical data makes it more flexible for intermittent demand forecasting, especially when the training set is limited in positive demands. In addition, the features in FIDE are all easily understood. The two features focusing on the presence of recent demand are proved more critical for constructing the forecast combination model.
These advantages of the proposed methods lead to broad application prospects in intermittent demand forecasting.

However, we recognize the lack of a comprehensive evaluation of inventory performance in the current study. \citet{petropoulos2019inventory} combined financial, operational, and service metrics to form a holistic measure for inventory control objectives.  \citet{ducharme2021forecasting} focused on stock-out events and proposed a novel metric called Next Time Under Safety Stock. The utility measures are essential to achieve a direct link between inventory holding costs and service levels in the production system.
Such analysis needs to proceed based on restocking policies, which are not available for the RAF and M5 datasets without any background information of inventory. Future research should investigate the inventory performance of our proposed framework in the field of a specific inventory management problem. Another limitation of this paper is lacking an automatic procedure to choose features for modeling FIDE. Several scholars have investigated selecting features automatically from a large number of features \citep{lubba2019catch22,theodorou2021exploring}. Although these approaches seem more general, they take over much computational time, and the selected features are often difficult to understand in the applications. Based on the results of our work, the nine features in FIDE are efficient and can be used as the benchmark pool of features for intermittent demand. In further research, we will study a standard procedure to select features automatically for the proposed framework, aiming to achieve both interpretability and computational efficiency.

\section*{Data Availability Statement}
The RAF dataset has been used in previous literature \citep{teunter2009forecasting,petropoulos2015forecast,kourentzes2021elucidate} and is available upon request.
The M5 competition \citep{makridakis2021m5} data involves the unit sales of 3049 products between 2011-01-29 and 2016-06-19 (1969 days). The first 1941 observations for model training can be obtained from \url{https://github.com/Mcompetitions/M5-methods}; the final 28 observations is available upon request.

\section*{Acknowledgments}

Yanfei Kang is supported by the National Natural Science Foundation of China
(No. 72171011). Feng Li is supported by the Beijing Universities Advanced Disciplines Initiative (No. GJJ2019163) and the Emerging Interdisciplinary Project of CUFE. This research was supported by Alibaba Group through the Alibaba Innovative Research Program and the high-performance computing (HPC) resources at Beihang University.

\section*{Declaration of Interest Statement}
No potential conflict of interest was reported by the authors.

\bibliographystyle{tfcad}
\bibliography{ref}

\end{document}